\documentclass[sigplan,screen,balance=false]{acmart}

\usepackage[export]{adjustbox}
\usepackage{booktabs}
\usepackage{caption}
\usepackage{datetime}
\usepackage{hyperref}
\usepackage{multirow}
\usepackage{subcaption}
\usepackage{tcolorbox}
\usepackage{algorithm}
\usepackage{tikz}
\usepackage{algpseudocode}
\usepackage{xcolor}

\usepackage{pifont}

\algrenewcommand\Call[2]{\textproc{#1}(#2)}

\setcopyright{cc}
\setcctype{by-nc-nd}
\acmDOI{10.1145/3814942.3816132}
\acmYear{2026}
\copyrightyear{2026}
\acmISBN{979-8-4007-2720-7/2026/06}
\acmConference[ISMM '26]{Proceedings of the 2026 ACM SIGPLAN International Symposium on Memory Management}{June 16, 2026}{Boulder, CO, USA}
\acmBooktitle{Proceedings of the 2026 ACM SIGPLAN International Symposium on Memory Management (ISMM '26), June 16, 2026, Boulder, CO, USA}
\acmSubmissionID{pldiws26ismmmain-p24-p}
\received{2026-04-03}
\received[accepted]{2026-05-04}

\title[Reconsidering ``Reconsidering Custom Memory Allocation'']{Reconsidering \texorpdfstring{\\}{} \texorpdfstring{``Reconsidering Custom Memory Allocation''}{"Reconsidering Custom Memory Allocation"}}



\author{Nicolas van Kempen}
\orcid{0000-0002-1708-0073}
\affiliation{
    \institution{University of Massachusetts Amherst}
    \city{Amherst}
    \state{MA}
    \country{USA}
}
\email{nvankempen@cs.umass.edu}

\author{Emery D. Berger}
\authornote{Work done at the University of Massachusetts Amherst.}
\orcid{0000-0002-3222-3271}
\affiliation{
    \institution{University of Massachusetts Amherst}
    \city{Amherst}
    \state{MA}
    \country{USA}
}
\affiliation{
    \institution{Amazon Web Services}
    \city{Seattle}
    \state{WA}
    \country{USA}
}
\email{emery@cs.umass.edu}

\begin{document}

\begin{abstract}
    Programmers using native languages such as C, C++, or Rust can implement custom memory allocation strategies to improve execution time.
In their paper titled ``Reconsidering Custom Memory Allocation'' almost 25 years ago, Berger et al.~\cite{10.1145/582419.582421}
showed that while per-class allocators provide no significant speedups over a state-of-the-art general-purpose allocator,
region-based allocators can improve execution time by allocating and freeing objects in bulk.
This paper revisits that work on a modern hardware platform with modern general-purpose allocators
to evaluate whether their conclusions still hold.
It also augments the benchmark suite with two large real-world applications (Clang and Blender), and introduces a methodology to explore the effect
of memory fragmentation on locality in general-purpose allocators.
Our results support and extend the original conclusions, demonstrating the locality advantages of region-based custom memory allocators.

\end{abstract}

\maketitle

\section{Introduction}

Low-level control over memory management is a fundamental aspect of native programming languages such as C, C++, or Rust.
While garbage-collected systems provide increased safety guarantees and other software engineering benefits, manual memory management can provide significant performance improvements~\cite{10.1145/1094811.1094836}.
Notably, these performance enhancements can be obtained using a \textit{custom memory allocator} by leveraging knowledge about the application's memory access patterns and object lifetimes.

Berger et al.~\cite{10.1145/582419.582421} tackled the question of custom allocator
performance over two decades ago in a paper entitled
``Reconsidering Custom Memory Allocation''\footnote{\href{https://sigplan.org/Awards/OOPSLA/}{Awarded Most Influential OOPSLA Paper in 2012.}}.
Their study classifies custom allocation techniques, and measures their performance across a suite of eight benchmarks.
Those prior experiments identify \textit{region allocators} as the only custom allocation technique providing substantial improvements to execution time (up to 44\%) compared to a state-of-the-art general-purpose allocator.
Region allocators (also named arena, pool, or bump-pointer allocators) operate by allocating objects contiguously without individually freeing them, instead returning all memory at once when the region is destroyed.
This scheme allows for reduced allocator pressure thanks to fewer individual allocation and free operations:
using knowledge of object lifetimes within the application, objects with identical or similar lifetimes can be grouped together, allowing efficient bulk interactions with the underlying allocator.
Today, region allocators are still widespread. Most notably, they have been incorporated as first-class members of the C++ language since C++17~\cite{P3002, P0339, P2126, P2127, P2035, N4468}.
Region allocator implementations are also available in many third-party libraries, including APR, Boost, and Folly for C/C++, as well as Bumpalo and typed-arena for Rust.

Over the two decades since Berger et al.'s study, hardware and state-of-the-art general-purpose allocators have changed substantially.
Modern general-purpose allocators such as jemalloc~\cite{jemalloc} and mimalloc~\cite{mimalloc} have significantly narrowed the performance gap with custom strategies.
At the same time, the growing gap between CPU speed and memory latency means cache misses are proportionally more expensive today~\cite{10.1145/216585.216588, 10.1145/977091.977115}, making locality of reference an increasingly critical factor.
Beyond the hardware and software landscape, we identify a methodological limitation in prior evaluations:
benchmarks are run on a fresh, unfragmented heap, which places the general-purpose allocator in ideal starting conditions.
During execution, programs typically build up some level of natural heap fragmentation from earlier operations within the same process.
Starting from a blank heap inadvertently advantages the general-purpose allocator, particularly for short-lived benchmarks.

To address this methodological gap, we introduce \textit{adversarial allocation},
a technique that preconditions the heap with artificial fragmentation before the benchmark runs.
Adversarial allocation first profiles the program to record its allocation size distribution
and peak live object count, then uses this information to allocate and randomly free a large number of objects,
leaving the heap in a fragmented state that mimics conditions in long-running or shared environments.
On this preconditioned heap, general-purpose allocators must fill gaps scattered across memory, whereas
region allocators still guarantee contiguous placement of objects.
This exposes a locality benefit of region allocators that clean-heap evaluations systematically obscure:
consecutive allocations within a region remain spatially co-located regardless of prior heap state,
reducing cache misses when those objects are later traversed.

This paper makes the following contributions:
\begin{itemize}
    \item Revisits custom allocators in the face of modern hardware and better high-performance general-purpose allocators.
    \item Proposes \textit{adversarial allocation}, a refined methodology for accurately quantifying the locality upside
    of region allocators, without inadvertently giving an advantage to na\"ive general-purpose allocation.
    \item Evaluates the impact of region allocators in real programs, tackling benchmarks from Berger et al.'s evaluation and from two large-scale modern applications.
\end{itemize}

Our results largely support and extend Berger et al.'s original conclusions:
per-class allocators provide no substantial benefit over a modern general-purpose allocator,
while region allocators retain a meaningful advantage, particularly under heap fragmentation.
However, our experiments also show that high-performance general-purpose allocators such as jemalloc
and mimalloc have considerably closed the performance gap with region-based custom allocation,
reducing region allocator speedups from up to 44\% in the original study to at most 15\% in our evaluation.
Adversarial allocation further reveals a resilience advantage of region allocators
that clean-heap evaluations miss entirely: na\"ive allocation slows down by up to $2\times$
under heap fragmentation, while region allocators are unaffected.

\section{Background}
\subsection{General-Purpose Memory Allocators}
\label{sec:gp-allocators}

Dynamic memory allocation is necessary in low-level languages such as C and C++.
General-purpose allocators must satisfy a wide set of requirements, without any knowledge
of the application's specific allocation patterns: low overhead per operation,
high throughput even in multi-threaded workloads, low memory waste, and good memory locality.
A key technique used by general-purpose allocators is \emph{size-class segregation}:
the heap is partitioned into pools of fixed-size slots, and each allocation
request is rounded up to the nearest size class and served from the
corresponding pool.

Berger et al.'s experiments focus on two general-purpose allocators:
the default Windows XP allocator (``Win32''), and the Doug Lea allocator (``dlmalloc'')~\cite{dlmalloc}.
At the time, dlmalloc was a state-of-the-art high-performance allocator. Today, both of those allocators are severely outdated.
This paper discusses three general-purpose allocators widely used today: glibc malloc, jemalloc~\cite{jemalloc}, and mimalloc~\cite{mimalloc}.
glibc malloc evolved from dlmalloc and is the default Linux allocator. jemalloc and mimalloc are modern high-performance allocators.

\subsection{Custom Allocators}
\label{sec:custom}

Using a custom allocator is a common technique to obtain performance gains and/or software engineering benefits.
Most often, the motivation for implementing a custom allocator is performance.
In some applications, custom allocators also allow for easier memory management;
for example, a server handling requests may use a custom allocator to facilitate memory reclamation of objects after request handling is complete.

Custom allocators are an old idea, but they are still used today. 
In fact, there has been significant work to integrate custom allocator support as an integral part of the C++ standard library with C++17~\cite{P3002, P0339, P2126, P2127, P2035, N4468}.
This integration is in the form of the \texttt{std::pmr} library, which provides a standard interface for custom allocators named \texttt{memory\_resource}.
A few subclasses of \texttt{memory\_resource} are provided in the standard library, most notably a region allocator implementation named \texttt{monotonic\_buffer\_resource}.

\begin{table*}[tp]
  \centering
  \caption{\textbf{Classification of the custom allocators used in Berger et al.'s original study and in this paper (\textsection{}\ref{sec:custom}).}
  Allocators are marked either per-class, region, or hybrid. This table also lists specific characteristics and interface details
  of those custom allocators: whether or not they use chunks, and what free interface is available.
  Three benchmarks from the original study are excluded: we could not find complete sources for C-Breeze, or a working and useful input for lcc.
  Apache's benchmarking is noisy and not compute-intensive enough to draw meaningful allocator performance conclusions.}
  \label{tab:benchmark-taxonomy}
  \begin{tabular}{llcccccl}
    \toprule
    Benchmark & Type & Original & This Study & Chunks & Individual Free & Bulk Free & Reason for Exclusion \\
    \midrule
    boxed-sim & per-class & \checkmark & \checkmark & & \checkmark & \\
    mudlle & region & \checkmark & \checkmark & \checkmark & & \checkmark & \\
    175.vpr & region & \checkmark & \checkmark & \checkmark & & \checkmark \\
    176.gcc & region & \checkmark & \checkmark & \checkmark & & \checkmark \\
    197.parser & hybrid & \checkmark & \checkmark & \checkmark & \checkmark & \\
    \midrule
    Clang & region & & \checkmark & \checkmark & & \checkmark \\
    geometry\_nodes & region & & \checkmark & \checkmark & & \checkmark \\
    sculpt & hybrid & & \checkmark & \checkmark & \checkmark & \checkmark \\
    \midrule
    C-Breeze & per-class & \checkmark & & & \checkmark & & missing sources \\
    Apache & region & \checkmark & & \checkmark & & \checkmark & not compute-bound \\
    lcc & region & \checkmark & & \checkmark & & \checkmark & missing benchmark input \\
    \bottomrule
  \end{tabular}
\end{table*}

We follow the taxonomy established and used by Berger et al.~\cite{10.1145/582419.582421}.
Namely, we split allocators into three categories: per-class, region, and \emph{hybrid}.
Table~\ref{tab:benchmark-taxonomy} provides an overview of the benchmarks used in this paper
and in the original study, and a few characteristics of their custom allocator.
The following sections provide more explanation of each type, along with examples taken from those benchmarks.

\subsubsection{Per-Class Allocators}
\label{section:per-class-explanation}

Per-class allocators optimize for a single object size/type, providing the usual \texttt{malloc}/\texttt{free} API.
On deallocation, they keep a freelist of objects ready to be reused on the next allocation instead of returning memory to the system allocator.
To minimize memory overhead, linked list pointers can be embedded in freed objects.
Section~\ref{sec:gp-allocators} describes how general-purpose allocators already separate their allocations into size classes:
therefore, intuitively, per-class allocators may only offer limited savings from size computations. Section~\ref{sec:rq1} quantifies these savings.

\subsubsection{Region Allocators}
\label{section:custom-explanation}

\algnewcommand{\algorithmicforeach}{\textbf{for each}}
\algnewcommand{\algorithmicin}{\textbf{in}}
\algrenewcommand\algorithmicdo{\textbf{do}}
\algblockdefx[ForEach]{ForEach}{EndForEach}[2]{\algorithmicforeach\ #1\ \algorithmicin\ #2\ \algorithmicdo}{\algorithmicend\ \algorithmicforeach}

\begin{algorithm}[tp]
\caption{\textbf{Region Allocation (\textsection{}\ref{section:custom-explanation}).} If the current chunk doesn't have sufficient space, allocate a new one and set it as the current chunk.
Then, bump the current chunk's allocation pointer. Allocators often have additional characteristics, such as alignment, special handling of larger allocations, 
or monotonically increasing chunk sizes.}
\label{alg:region-allocate}
\begin{algorithmic}
\Function{AllocateObject}{region, size}
    \If{\Call{Available}{region.current\_chunk} < size}
        \State region.current\_chunk $\gets$ \Call{CreateNewChunk}{}
        \State \Call{Append}{region.chunks, region.current\_chunk}
    \EndIf
    \State object $\gets$ \Call{CurrentBumpPtr}{region.current\_chunk}
    \State \Call{IncrementBumpPtr}{region.current\_chunk, size}
    \State \Return object
\EndFunction
\end{algorithmic}
\end{algorithm}

\begin{algorithm}[tp]
\caption{\textbf{Region Reset (\textsection{}\ref{section:custom-explanation}).} In region allocators, memory is retained until the entire region is ready to be disposed of.
This is one source of savings region allocators provide: significantly reduced total memory operations.}
\label{alg:region-reset}
\begin{algorithmic}
\Function{Reset}{region}
    \ForEach{chunk}{region.chunks}
        \State free(chunk)
    \EndForEach
\EndFunction
\end{algorithmic}
\end{algorithm}

Region allocators~\cite{10.1145/277650.277748} aim to group together objects that share a similar lifetime and/or will be frequently accessed together.
They follow the idea of a ``bump-pointer''; incrementing a pointer for each object allocation, freeing everything in one single operation, but only
when finished using all allocated objects.
While some implementations use a fixed-size maximum, most others extend this functionality to a monotonically growing set of chunks,
allowing support for an arbitrary number of objects.
Algorithms~\ref{alg:region-allocate} and~\ref{alg:region-reset} provide the outline for a simple region allocator implementation.
On allocation, the pointer is simply bumped if the current chunk has enough available bytes, otherwise a new chunk must first be allocated.
On reset and destruction, all allocated chunks are simply passed directly to the free function.
This interface and implementation match C++17's \texttt{std::pmr::monotonic\_buffer\_resource}, both in libc++ and libstdc++.

Custom allocators in mudlle, 175.vpr, 176.gcc, and Clang, as well as Blender's LinearAllocator (geometry\_nodes),
all follow this mechanism. 176.gcc adds one special feature: the ability to partially free the region. By providing
a pointer, all memory from that pointer to the region's tail will be deallocated.

Region allocators may suffer from increased memory footprint due to deallocations being deferred until the entire region is disposed of.
This increased memory requirement is largely dependent on program and allocator implementation rather than measurement platform or machine architecture,
and has been studied by Berger et al.~\cite{10.1145/582419.582421}; this paper does not further discuss it.

\subsubsection{Hybrid Allocators}
\label{section:hybrid-explanation}

Allocators not strictly falling into either one of the categories above typically employ a combination of techniques.
In our benchmarks, this is the case for 197.parser and Blender's mempool (sculpt).

197.parser's allocator employs a fixed-size chunk of memory provisioned at startup. It then operates with a standard malloc/free API.
Freeing an object marks it as free, and if it is the last allocated object, the allocator resets its internal bump-pointer to the new last live object.
In other words, an object will be re-used only if all subsequently allocated objects are first marked free. This method is effective for a stack-like use of memory.

Blender's mempool mixes region-based and freelist-based custom allocation.
Each pool has a fixed object size, and a fixed number of objects per chunk, with new chunks allocated as needed from the system allocator.
In addition to standard region allocation, BLI\_mempool also has a freelist mechanism: on deallocation, objects
are kept in a linked list for immediate reuse on the next allocation (last in, first out).
Another particularity of this allocator is support for iteration over all allocated objects. During iteration,
objects continue to be allocated in and freed from the pool.

\section{Adversarial Allocation}
\label{sec:littering}
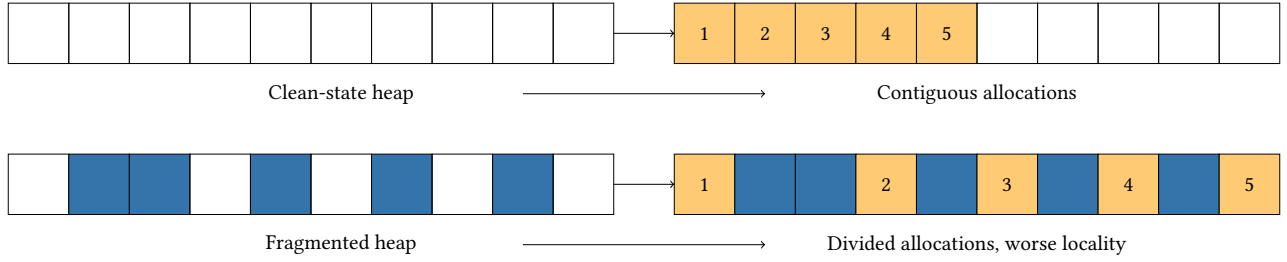
\begin{figure*}[tp]
    \centering
    \scalebox{0.8}{\begin{tikzpicture}[
            box/.style={
                rectangle,
                draw=black,
                minimum width=1cm,
                minimum height=1cm
            },
        ]

        \definecolor{myblue}{RGB}{2, 81, 150}
        \definecolor{myorange}{RGB}{253, 179, 56}

        \colorlet{garbage}{myblue!80}
        \colorlet{objects}{myorange!70}

        \node[box] at (0,0) {};
        \node[box] at (1,0) {};
        \node[box] at (2,0) {};
        \node[box] at (3,0) {};
        \node[box] at (4,0) {};
        \node[box] at (5,0) {};
        \node[box] at (6,0) {};
        \node[box] at (7,0) {};
        \node[box] at (8,0) {};
        \node[box] at (9,0) {};
        \node at (5,-1) {Clean-state heap};

        \node[box, fill=objects] at (11,0) {1};
        \node[box, fill=objects] at (12,0) {2};
        \node[box, fill=objects] at (13,0) {3};
        \node[box, fill=objects] at (14,0) {4};
        \node[box, fill=objects] at (15,0) {5};
        \node[box] at (16,0) {};
        \node[box] at (17,0) {};
        \node[box] at (18,0) {};
        \node[box] at (19,0) {};
        \node[box] at (20,0) {};
        \node at (15.5,-1) {Contiguous allocations};

        \draw[->] (9.5, 0) -- (10.5, 0);
        \draw[->] (8, -1) -- (12, -1);

        \node[box] at (0,-2.5) {};
        \node[box, fill=garbage] at (1,-2.5) {};
        \node[box, fill=garbage] at (2,-2.5) {};
        \node[box] at (3,-2.5) {};
        \node[box, fill=garbage] at (4,-2.5) {};
        \node[box] at (5,-2.5) {};
        \node[box, fill=garbage] at (6,-2.5) {};
        \node[box] at (7,-2.5) {};
        \node[box, fill=garbage] at (8,-2.5) {};
        \node[box] at (9,-2.5) {};
        \node at (5,-3.5) {Fragmented heap};

        \node[box, fill=objects] at (11,-2.5) {1};
        \node[box, fill=garbage] at (12,-2.5) {};
        \node[box, fill=garbage] at (13,-2.5) {};
        \node[box, fill=objects] at (14,-2.5) {2};
        \node[box, fill=garbage] at (15,-2.5) {};
        \node[box, fill=objects] at (16,-2.5) {3};
        \node[box, fill=garbage] at (17,-2.5) {};
        \node[box, fill=objects] at (18,-2.5) {4};
        \node[box, fill=garbage] at (19,-2.5) {};
        \node[box, fill=objects] at (20,-2.5) {5};
        \node at (15.5,-3.5) {Divided allocations, worse locality};

        \draw[->] (9.5, -2.5) -- (10.5, -2.5);
        \draw[->] (8, -3.5) -- (12, -3.5);
    \end{tikzpicture}}
    \caption{\textbf{Comparing the performance of short-lived benchmarks from a
    clean-state heap masks potential locality improvements brought by the region allocator (\textsection{}\ref{sec:littering}).}
    When starting from a blank, clean-state heap, the program's allocator can typically deliver
    contiguous allocations. A region allocator on the other hand guarantees this contiguity, even
    in the presence of heap fragmentation.}
    \label{fig:blank-heap-advantage}
    \Description{
        A stacked diagram comparing two memory layouts.

        Top row: A long horizontal bar represents a "clean-state heap", shown as evenly sized empty blocks.
        An arrow points to the right, where five adjacent blocks are highlighted and labeled 1 through 5.
        These highlighted blocks/objects sit together in one uninterrupted sequence, illustrating contiguous allocations.

        Bottom row: Another horizontal bar represents a "fragmented heap" where some blocks are filled (colored)
        and others are empty, scattered irregularly. An arrow points to the right, where five labeled blocks
        (1 through 5) appear separated by gaps and interleaved with other filled blocks.
        The same set of data is split across non-adjacent memory locations, reducing efficiency.
    }
\end{figure*}

Applications often gradually build up heap fragmentation throughout their execution, as allocations and deallocations of different sizes interleave.
Therefore, running short-lived benchmarks from a clean-state heap does not give the full picture of the benefits of custom allocators,
as general-purpose allocators generally behave optimally from a blank starting point, allocating contiguous objects.
Figure~\ref{fig:blank-heap-advantage} illustrates the effect of allocating from a blank heap versus an already fragmented heap:
in the blank heap case, memory allocators typically return objects in contiguous memory. In contrast, when the heap is fragmented,
memory allocators must fill distant gaps to keep memory footprint under control.

In order to fairly measure the full performance benefits of custom allocators, we introduce \textit{adversarial allocation},
a method preconditioning the heap to simulate heap fragmentation that may occur in long-running applications or in tail-latency cases.
Adversarial allocation consists of exercising the memory allocator before the program runs, simulating prior executions.
This technique requires two phases: first, the program runs with lightweight instrumentation to record the size class distribution of all allocations made.
This distribution is then used for all subsequent runs to repeatedly call malloc and free before the program's execution actually starts.

\begin{figure}[tp]
    \centering
    \includegraphics[width=0.99\linewidth]{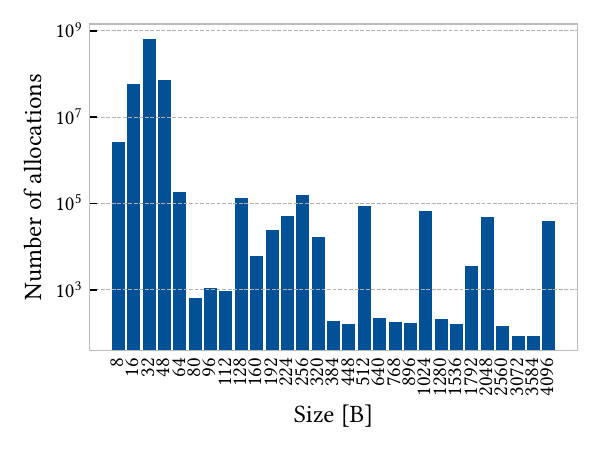}
    \caption{\textbf{Distribution of allocations for 197.parser, grouped in size class bins, with a log-scale (\textsection{}\ref{sec:adv-p1}).}
    The detection phase also tracks the maximum count of live allocations: 491k for this benchmark.}
    \label{fig:distribution-example}
    \Description{
        A bar chart showing how often different memory allocation sizes occur.

        Along the horizontal axis are allocation sizes in bytes,
        starting at 8 and increasing in steps up to 4096. Along the vertical axis is the number of allocations,
        marked in powers of ten.

        Each size corresponds to a vertical bar.
        Smaller sizes, especially between 8 and about 64 bytes, have the tallest bars, 
        indicating they are used more frequently. The tallest bar is at 32 bytes, with a value close to 1G allocations.
        As the sizes increase, the bars generally become shorter.
        There are occasional small bumps at certain size boundaries, such as 512, 1024, or 2048 bytes.
    }
\end{figure}

\subsection{Phase 1: Detecting Allocation Sizes}
\label{sec:adv-p1}

General-purpose allocators typically segregate allocations by size classes. Therefore, for representative artificial heap fragmentation,
we need to gather data on the program's allocation sizes.
We achieve this with a small library interposing on the standard memory allocation functions.
The output of this tool consists of the count of allocations for each size under 4096 bytes by default;
allocations of 4096 or more bytes are ignored as they meet or exceed the standard page size.
While this paper's benchmarks are not particularly impacted by physical memory footprint and address translation,
this cutoff setting can easily be adjusted to account for large allocations in applications sensitive to TLB misses.
Extending this study to huge pages is an interesting future direction.
Figure~\ref{fig:distribution-example} depicts an example allocation size distribution generated by this tool for 197.parser.
The tool also increments and decrements a counter on malloc and free respectively, recording the maximum number of live allocations
at any point in the program. Together with the distribution of allocation sizes, the adversarial allocation phase
will use this maximum live allocation number to obtain an estimated peak allocation footprint.

\begin{algorithm}[tp]
\caption{\textbf{Preconditioning the heap before program execution by allocating and randomly freeing objects (\textsection{}\ref{sec:precondition}).}
The size distribution and the peak number of live allocations are retrieved from the previous detection phase. The multiplier and
occupancy settings control the preconditioning footprint and density, respectively.}
\label{alg:littering}
\begin{algorithmic}
\Require distribution, peak\_allocations
\Function{PreCondition}{multiplier, occupancy}
    \State objects $\gets$ \Call{CreateEmptyArray}{}
    \State $n \gets \textnormal{multiplier} \times \textnormal{peak\_allocations}$
    \For{$i \in \left[ 1 ; n \right]$}
        \State size $\gets$ \Call{SampleRandomSize}{distribution}
        \State \Call{Append}{objects, \Call{Malloc}{size}}
    \EndFor
    \State \Call{Shuffle}{objects}
    \State $m \gets n \times \left(1 - \textnormal{occupancy}\right)$
    \For{$i \in \left[ 1 ; m \right]$}
        \State \Call{Free}{objects[i]}
    \EndFor
\EndFunction
\end{algorithmic}
\end{algorithm}

\subsection{Phase 2: Preconditioning the Heap}
\label{sec:precondition}

Algorithm~\ref{alg:littering} formalizes the preconditioning procedure happening during static initialization, before the program executes.
The allocation size distribution and the peak number of live allocations are retrieved from the previous detection phase.
With a given multiplier $M$, the tool allocates $M \times \texttt{peak\_allocations}$ objects,
sampling each object's size proportionally from the recorded distribution.
The multiplier simulates the possible heap state of a long-running application that has, over its lifetime,
allocated and mostly freed a quantity of memory far exceeding its current working set.
After allocation, the objects are randomly and uniformly shuffled, and freed, leaving
only a given occupancy fraction of them permanently live on the heap.
By randomizing the order of allocated objects before freeing, the algorithm ensures that live and free blocks are interleaved
throughout the heap, producing a fragmented free list that resembles what accumulates in practice~\cite{DBLP:conf/iwmm/WilsonJNB95}.

\begin{figure}[tp]
    \centering
    \begin{subfigure}{\linewidth}
        \centering
        \includegraphics[width=0.9\columnwidth]{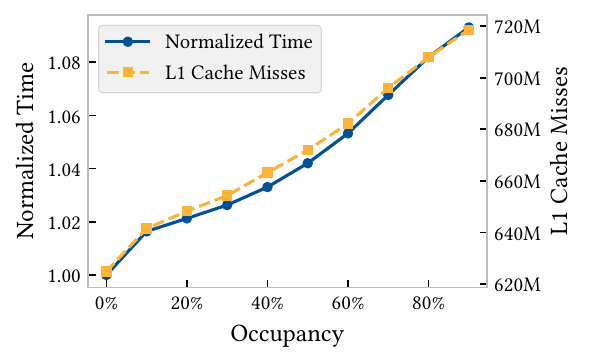}
        \caption{Clang with jemalloc}
    \end{subfigure}
    \begin{subfigure}{\linewidth}
        \centering
        \includegraphics[width=0.9\columnwidth]{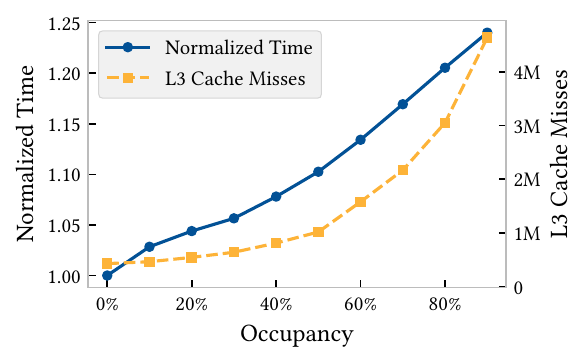}
        \caption{197.parser with jemalloc}
    \end{subfigure}
    \caption{\textbf{Adversarial allocation occupancy controls synthetic heap fragmentation (\textsection{}\ref{sec:precondition}).}
    These two examples, covering L1 misses in Clang and L3 misses in 197.parser under
    a fixed adversarial allocation multiplier of 10,
    highlight the correlation between cache misses (poor locality) and increased execution time.}
    \label{fig:cycle}
    \Description{
        Two stacked line charts that show higher adversarial allocation occupancy leads to more cache misses and slower execution.

        For both charts:
        The horizontal axis is adversarial allocation occupancy, ranging from 0\% to 90\%.
        The left vertical axis shows normalized execution time, from 1 to 1.10 for the top chart, 1 to 1.25 for the bottom chart.
        The right vertical axis shows cache misses, L1 from 620M to 720M for the top chart, L3 from 0 to 5M for the bottom chart.
        Two lines are plotted: a solid blue line for normalized time and a dashed orange line for cache misses.
        Both lines rise steadily as occupancy increases.
    }
\end{figure}

Performance counters show that adversarial allocation increases
cache misses at every level, indicating that heap fragmentation is the
primary cause of the observed slowdown.
Figure~\ref{fig:cycle} shows examples of this effect, by gradually increasing adversarial allocation occupancy with a fixed multiplier.
As occupancy increases, fewer contiguous object slots remain available to the allocator during execution.
In turn, cache misses increase at every level, degrading performance.
Adversarial allocation is a useful tool to quantify a program's resilience to memory fragmentation,
even when it is not always representative of realistic long-term fragmentation occurring in a given program.
General-purpose allocators may implement heuristics to defeat this particular implementation, but an adversarial pattern can always be constructed for any allocator~\cite{DBLP:conf/iwmm/WilsonJNB95}.

\section{Evaluation}
\label{sec:evaluation}
\begin{table*}[tp]
  \centering
  \caption{\textbf{Applications used to evaluate custom allocators, along with their version and input (\textsection{}\ref{sec:evaluation}).}}
  \label{tab:benchmarks}
  \begin{tabular}{llll}
    \toprule
    Benchmark & Version & Input & Description \\
    \midrule
    boxed-sim & sha256:fc65ed0 & -n 128 -s 13 & simulates $n$ balls bouncing inside a box\\
    mudlle & 2020-04-30 & time.mud & MUME's extension language interpreter\\
    175.vpr & spec2000v1.3 & train input & FPGA circuit placement and routing \\
    176.gcc & spec2000v1.3 & 200.i & C compiler \\
    197.parser & spec2000v1.3 & ref input & natural language processing \\
    Clang & 21.1.8 & sqlite3.c & C/C++ compiler \\
    geometry\_nodes & 5.0.1 & foreach\_zone\_bfield & geometry node modification \\
    sculpt & 5.0.1 & $1000\times{}1000$ mesh & diagonal brush stroke on mesh \\
    \bottomrule
  \end{tabular}
\end{table*}

\begin{table}[tp]
  \centering
  \caption{\textbf{Configurations of adversarial allocations used in our evaluation (\textsection{}\ref{sec:evaluation}).}
    Each configuration is further tested with the region allocator enabled or falling back to the underlying allocator,
    and with three backing general-purpose allocators: glibc malloc, jemalloc, and mimalloc.}
  \label{tab:adv-configs}
  \begin{tabular}{lrr}
    \toprule
    Name & Multiplier & Occupancy \\
    \midrule
    $Adv_0$    & 0  & --   \\
    $Adv_1$    & 1  & 0.33 \\
    $Adv_3$    & 3  & 0.66 \\
    $Adv_{10}$ & 10 & 0.8  \\
    \bottomrule
  \end{tabular}
\end{table}

\newcommand{\rqone}{\textbf{RQ1}: Do per-class custom allocators provide meaningful performance benefits?}
\newcommand{\rqtwo}{\textbf{RQ2}: Have high-performance general-purpose allocators closed the gap with region allocators?}
\newcommand{\rqthree}{\textbf{RQ3}: Do region allocators offer more resilience to heap fragmentation?}

Berger et al.'s experimental setup used a ``600 MHz Pentium III system with 320MB of RAM, a unified 256K L2 cache,
and 16K L1 data and instruction caches''. For reference, Pentium III processors were discontinued in 2004.
In contrast, this paper's evaluation was conducted on a server equipped with two Intel Xeon Gold 6430 processors (initial release: 2023), running Linux version 6.8.0-84-generic, with 128GB of memory and a total of 120MB L3 cache.
The tested underlying general-purpose allocators are glibc malloc 2.39, jemalloc 5.3.0, and mimalloc 2.2.5.
We compile both benchmarks and allocators with Clang 21 and all optimizations enabled, except for glibc malloc which
comes from the system's library. Reported results are averaged over repeated runs; variance was low across all measurements.
Table~\ref{tab:adv-configs} details all adversarial allocation settings discussed in our evaluation.
This section aims to address the following research questions:
\begin{itemize}
  \item \rqone
  \item \rqtwo
  \item \rqthree
\end{itemize}
The first two research questions mirror and aim to refresh Berger et al.'s evaluation.
The third question extends our experiments to adversarial allocation and resilience to heap fragmentation.

\subsection{Benchmarks}
\label{sec:benchmarks}

Table~\ref{tab:benchmarks} describes the benchmarks used, as well as their version and input.
We include all benchmarks from Berger et al.'s study we could successfully obtain and run,
namely boxed-sim, mudlle, 175.vpr, 176.gcc, and 197.parser~\cite{speccpu2000}.
To obtain modern real-world examples of C and C++ programs using custom allocation, we searched Google and GitHub, and queried the
ChatGPT and Claude large language models for projects using custom allocator implementations.
Candidate applications were then filtered, selecting those amenable to meaningful benchmarking of their custom allocator.
This process allows us to extend our evaluation with three additional benchmarks representative of modern application workloads:
one from LLVM/Clang~\cite{clang} and two from Blender~\cite{blender}.

Clang~\cite{clang} is a very widely used C and C++ compiler, based on the LLVM toolchain~\cite{10.5555/977395.977673}.
Among a few other less frequently used custom allocator strategies, it implements a region allocator in \texttt{llvm::BumpPtrAllocator}.
This implementation follows the general description in Section~\ref{section:custom-explanation}:
it provides \textsc{Allocate} and \textsc{Reset} methods that respectively allocate an object of the given size, and free all allocated objects.
We benchmark Clang on the SQLite3 v3.49.1 amalgamation, a very large (165k lines of code) C file, with all optimizations enabled (\texttt{-O3}).

Blender~\cite{blender} is a state-of-the-art 3D computer graphics tool, widely used for 3D modeling and animations, including full-length animated films~\cite{flow}.
Its source contains three different implementations of region allocators, each used in different subsystems.
This evaluation focuses on two:
\begin{enumerate}
  \item LinearAllocator: This allocator follows the general mechanism outlined in Section~\ref{section:custom-explanation}.
  \item BLI\_mempool: This hybrid allocator mixes freelist-based and region-based custom allocation, and supports iteration.
    Notably, it requires stability during iteration in the event of object allocations or frees.
\end{enumerate}
We benchmark those two allocators with built-in Blender performance benchmarks, with some changes to input size and/or output format:
geometry\_nodes for LinearAllocator and sculpt for BLI\_mempool.

\subsection{Disabling Custom Allocation}
\label{sec:disable-custom}

\begin{table}[tp]
  \centering
  \caption{\textbf{Methods to disable custom allocation (\textsection{}\ref{sec:disable-custom}).}}
  \label{tab:disable-custom}
  \begin{tabular}{ll}
    \toprule
    Benchmark & Disable Method \\
    \midrule
    boxed-sim & remove freelist \\
    mudlle & reduce chunk size to fit single objects \\
    175.vpr & reduce chunk size to fit single objects \\
    176.gcc & reduce chunk size to fit single objects \\
    197.parser & use malloc/free directly \\
    Clang & reduce large allocation threshold to 0  \\
    geometry\_nodes & reduce large allocation threshold to 0 \\
    sculpt & track live objects with a hash set \\
    \bottomrule
  \end{tabular}
\end{table}

To effectively compare custom allocation strategies to relying solely on the general-purpose allocator,
we maintain two separate versions of each benchmark:
a first version as-is, with the custom allocator enabled, and a second where all individual object requests are forwarded directly into the system's malloc and free functions (na\"ive allocation). 
We use the same methodology as Berger et al., and manually replace each custom allocator so that it falls back to the default underlying system
allocator in the most efficient way possible.
Disabling custom allocation can be as simple as redirecting back to the underlying malloc/free functions (197.parser).
For most region allocators, ensuring each chunk contains exactly one allocation is a simple but effective way to guarantee objects are allocated via malloc, while maintaining object ownership tracking and bulk free capabilities.
Per-class allocators (boxed-sim) are also straightforward to replace by removing each type-specific freelist.
Blender's mempool is the most complex allocator to disable, due to its wide interface features:
it must support individual and bulk frees, as well as iteration over all allocated objects.
In this case, the most efficient way to emulate na\"ive allocation
is keeping a hash set of all live objects.
Table~\ref{tab:disable-custom} summarizes the methods employed to disable custom allocation in each benchmark.

\subsection{\rqone}
\label{sec:rq1}

\begin{figure*}[tp]
    \centering
    \includegraphics[scale=0.7]{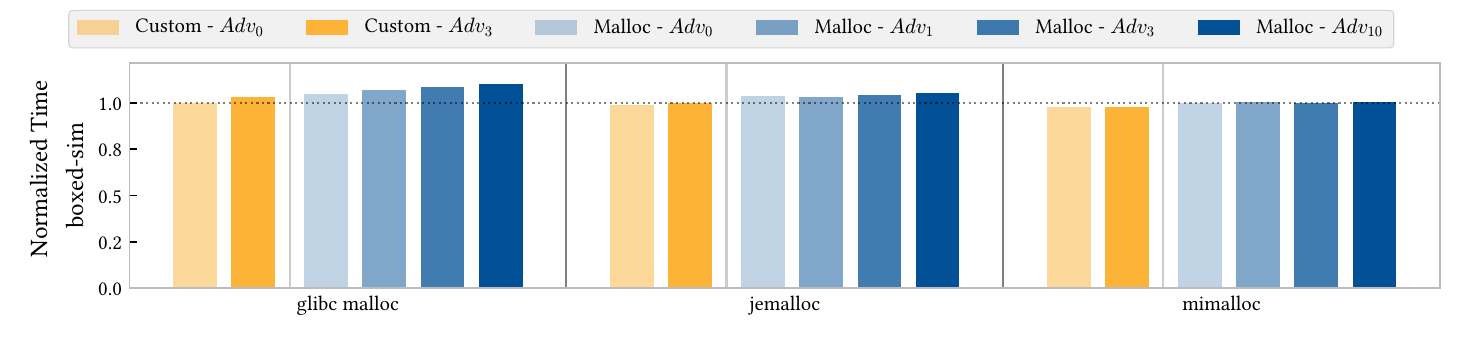}
    \caption{\textbf{Per-class custom allocation in boxed-sim only provides marginal execution time improvements (\textsection{}\ref{sec:rq1}).}
      Execution time is normalized to using the custom allocator under glibc malloc with no adversarial allocation. Table~\ref{tab:adv-configs} describes the adversarial allocation settings used.}
    \label{fig:boxed-sim}
    \Description{
        A grouped bar chart displaying normalized execution time for boxed-sim under various configurations.

        The y-axis is normalized execution time. The x-axis is split into three groups: glibc malloc, jemalloc, and mimalloc.
        For each allocator group, there are 6 bars: Custom - $Adv_0$, Custom - $Adv_3$, Malloc - $Adv_0$, Malloc - $Adv_1$, Malloc - $Adv_3$, and Malloc - $Adv_{10}$.
        All bars are near 1, showing no significant performance differences.
    }
\end{figure*}

This section focuses on the boxed-sim benchmark. It uses a per-class allocator:
on deallocation, objects are placed on class-local freelists and are recycled for the next allocation (last in, first out).
Figure~\ref{fig:boxed-sim} shows normalized execution time results for this benchmark.
For boxed-sim, na\"ive allocation with mimalloc outperforms glibc malloc with the custom allocator even on the highest adversarial allocation settings.
Custom per-class allocation on top of mimalloc only shaves an additional 2.3\%.

Berger et al.~discuss one additional benchmark using per-class allocators: C-Breeze.
However, it has not received any updates in two decades, and we were unable to successfully compile and run the program on our experimental platform.
Further, our search for modern C and C++ applications meaningfully using per-class allocation yielded no compelling candidates.

Nevertheless, we extend our experiments with an upper-bound microbenchmark:
it exclusively measures allocation throughput using intrusive pointers for
zero memory overhead in building the linked freelist.
With 16-byte allocations, this upper bound shows the per-class allocator
outperforms na\"ive allocation with glibc malloc by only 24\%,
jemalloc by 11\%, and mimalloc by 8\%.
In a realistic application where memory operations account for only a fraction
of total cycles, these gains drop by an order of magnitude.

Per-class allocators offer no additional protection against heap fragmentation.
The freelist ordering depends on deallocation history:
even if objects are initially allocated contiguously,
they become arbitrarily interleaved as allocations and frees occur.

\begin{tcolorbox}[colback=green!5!white,colframe=green!75!black, arc=1mm, top=0.5mm, bottom=0.5mm, left=0.5mm, right=0.5mm, enlarge top initially by=1mm]
\textbf{RQ1 Summary:}
The boxed-sim per-class allocator provides only a
2.3\% improvement to execution time with mimalloc, while requiring significant implementation effort.
The per-class allocation microbenchmark confirms that potential savings are minimal.
These results suggest that, in most cases, the engineering complexity of per-class custom
allocators outweighs their performance benefits, which modern general-purpose allocators already largely eliminate.
This aligns with Berger et al.'s conclusions.
\end{tcolorbox}

\subsection{\rqtwo}
\label{sec:rq2}

\begin{figure*}[tp]
    \centering
    \includegraphics[scale=0.7]{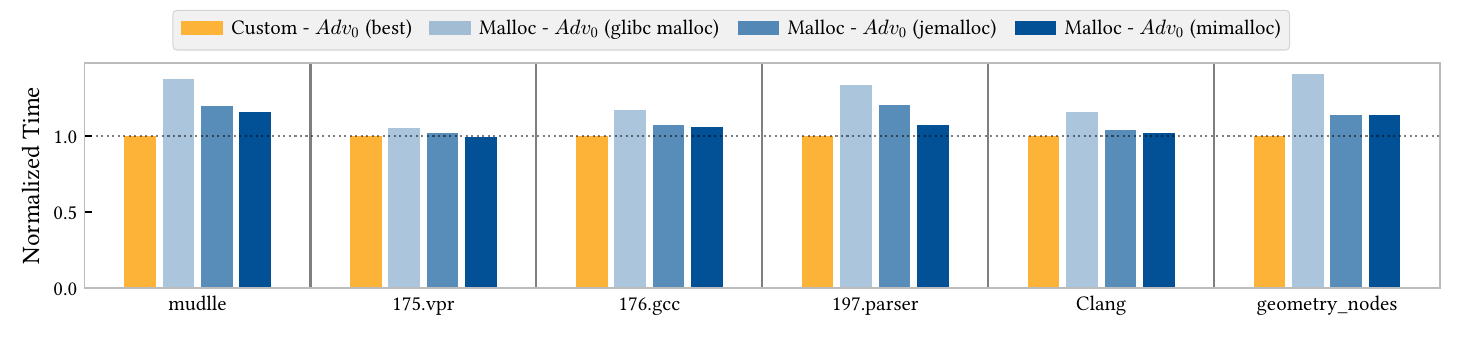}
    \caption{\textbf{Region allocators significantly outperform na\"ive general-purpose allocation in four benchmarks (\textsection{}\ref{sec:rq2}).}
      For each benchmark, execution time is normalized to using custom allocation with the best underlying general-purpose allocator.
      While Section~\ref{sec:custom} classifies 197.parser as using a hybrid allocator,
      its use of chunked allocation exhibits roughly the same reduction in memory operations as other region allocator benchmarks.}
    \label{fig:regions-no-adv}
    \Description{
        A grouped bar chart displaying normalized execution time for 6 benchmarks.

        The y-axis is normalized execution time. The x-axis is split into 6 groups: mudlle, 175.vpr, 176.gcc, 197.parser, clang, and geometry\_nodes.
        For each benchmark, there are 4 bars: Custom - $Adv_0$ (best), Malloc - $Adv_0$ (glibc malloc), Malloc - $Adv_0$ (jemalloc), and Malloc - $Adv_0$ (mimalloc).
        mudlle, 176.gcc, 197.parser, and geometry\_nodes all show better performance with the custom allocator rather than na\"ive allocation with any of the allocators.
    }
\end{figure*}

Figure~\ref{fig:regions-no-adv} summarizes and compares the execution time using custom allocation
(with the best possible underlying general-purpose allocator) versus na\"ive allocation.
Since no adversarial allocation is involved, execution time reductions can for the most part be attributed to the reduction
in memory operations that chunked allocators provide.
Berger et al.~\cite{10.1145/582419.582421} reported that ``region-based allocators often outperform general-purpose allocation'';
our results suggest high-performance general-purpose allocators today have significantly closed that gap.
Among the three tested general-purpose allocators, mimalloc provides the best overall performance on our selection of benchmarks.
Na\"ive allocation with mimalloc provides equal or nearly equal performance to custom allocation for 175.vpr and Clang.
For 197.parser, geometry\_nodes, and mudlle, the slowdown is 7.1\%, 14\%, and 15\%, respectively.

\begin{tcolorbox}[colback=green!5!white,colframe=green!75!black, arc=1mm, top=0.5mm, bottom=0.5mm, left=0.5mm, right=0.5mm, enlarge top initially by=1mm]
\textbf{RQ2 Summary:}
High-performance general-purpose allocators have significantly closed the performance gap with region allocators,
down from the 44\% upper bound reported by Berger et al.
Regions can still offer execution time improvements, up to 15\% in mudlle.
\end{tcolorbox}

\subsection{\rqthree}
\label{sec:rq3}

\begin{figure*}[p]
    \centering
    \includegraphics[scale=0.66]{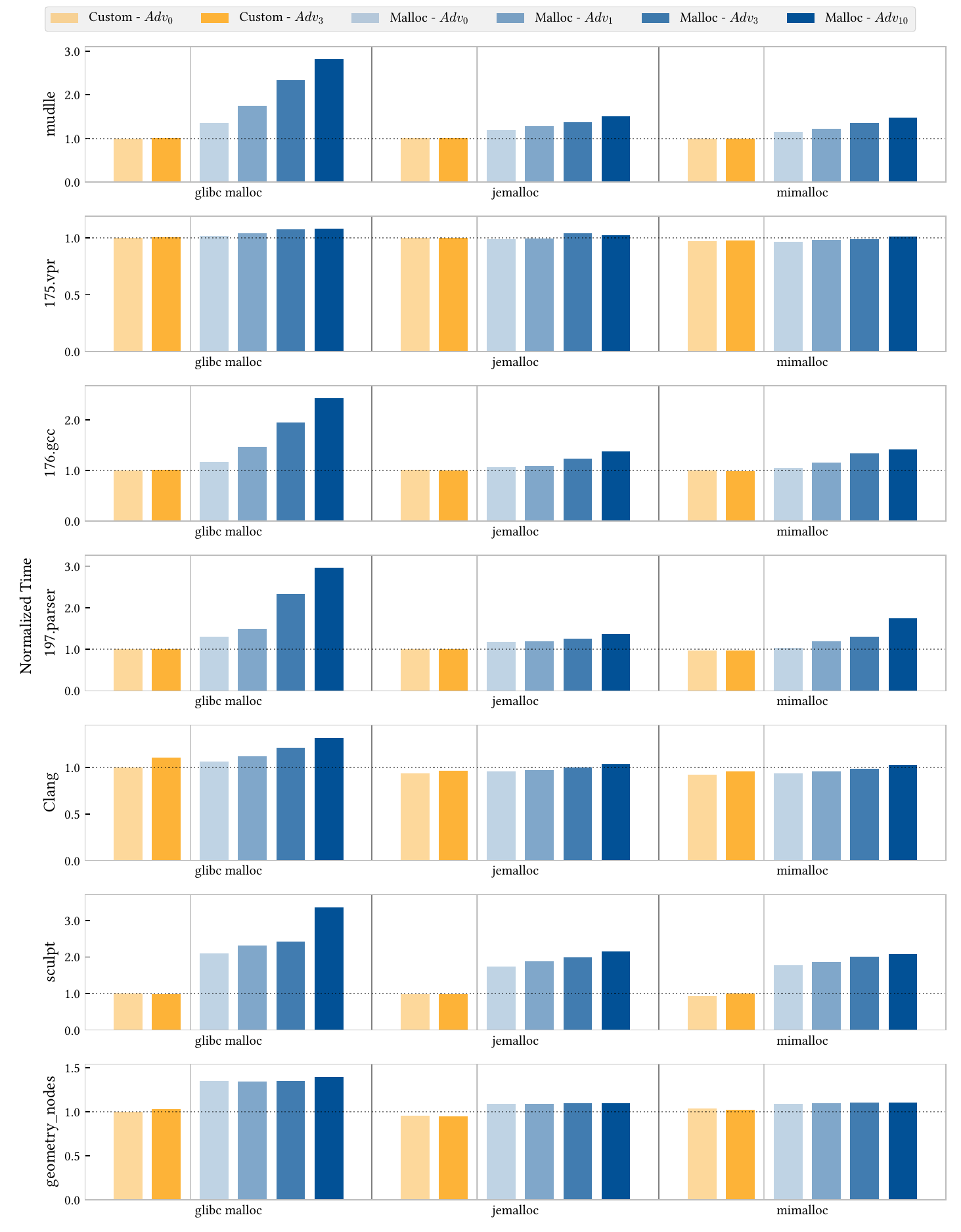}
    \caption{\textbf{Region allocators offer significant resilience to adversarial allocation, and more generally heap fragmentation (\textsection{}\ref{sec:rq3}).}
      For each benchmark, execution time is normalized to using the custom allocator under glibc malloc with no adversarial allocation.
      Table~\ref{tab:adv-configs} describes the adversarial allocation settings used.}
    \label{fig:all-runtime}
    \Description{
        7 stacked grouped bar charts displaying normalized execution time for 7 benchmarks, one per row: mudlle, 175.vpr, 176.gcc, 197.parser, clang, geometry\_nodes, and sculpt.

        For each row/benchmark, the bar chart is split into three groups: glibc malloc, jemalloc, and mimalloc.
        For each row and for each allocator, there are 6 bars: Custom - $Adv_0$, Custom - $Adv_3$, Malloc - $Adv_0$, Malloc - $Adv_1$, Malloc - $Adv_3$, and Malloc - $Adv_{10}$.
        Generally, benchmarks get worse as adversarial allocation multiplier and occupancy increase.
        This effect is very prominent with glibc malloc, and less so with jemalloc or mimalloc.

        This figure highlights:
        1. Custom allocation outperforms na\"ive allocation, even with high-performance modern allocators, on some benchmarks.
        2. Custom allocation is immune to heap fragmentation, while na\"ive allocation performance decreases with each successively worse adversarial allocation setting.
    }
\end{figure*}

This section leverages adversarial allocation, described in Section~\ref{sec:littering},
to better surface any potential locality-related benefits region allocators can bring.
Region allocators are able to fully bypass any natural fragmentation that gradually builds up
during program execution, making them ideal candidates in tail-latency and locality-sensitive workloads.

Figure~\ref{fig:all-runtime} shows the runtime for all region-allocator-based benchmarks in this paper.
Nearly every region allocator --- including the hybrid allocators in 197.parser and sculpt, whose chunked allocation provides the same locality properties --- shows perfect resilience to adversarial allocation, even with higher settings.
The only exception is Clang: this is because not all allocations in Clang go through the region allocator.
Instead, regions are integrated only in high-impact functions and procedures throughout the program.
Therefore, the remaining allocations are still susceptible to performance degradation from heap fragmentation.
Nevertheless, under $Adv_3$ settings, Clang with its region allocator still outperforms na\"ive allocation by
3.4\% for jemalloc, 3.2\% for mimalloc, and 9.4\% for glibc malloc.

In contrast, na\"ive allocation is heavily impacted by adversarial allocation, even
with modern high-performance allocators.
Based on the $Adv_3$ and mimalloc settings, adversarial allocation slows down
mudlle, 197.parser, and 176.gcc by 18\%, 25\%, and 27\%, respectively.
This slowdown is mitigated using custom, region-based allocation.
The default Linux allocator is considerably more affected by fragmentation than jemalloc and mimalloc,
with both mudlle and 197.parser taking over $2\times$ longer under $Adv_3$ settings.

Interestingly, 175.vpr and geometry\_nodes appear relatively unaffected by adversarial allocation.
175.vpr is very processor-intensive with a low memory footprint, and hence has low sensitivity to fragmentation issues.
70\% of allocations in the geometry\_nodes benchmark are bigger than 64 bytes, the cache line size on our experimental platform.
It also has a small working set of objects: only 18k.
Because of their allocation properties, locality in these programs is not a major concern;
region allocation can still provide performance benefits from reduced memory operations (Section~\ref{sec:rq2}).

\begin{tcolorbox}[colback=green!5!white,colframe=green!75!black, arc=1mm, top=0.5mm, bottom=0.5mm, left=0.5mm, right=0.5mm, enlarge top initially by=1mm]
\textbf{RQ3 Summary:}
Region allocators provide strong resilience to heap fragmentation,
while na\"ive allocation degrades execution time by up to $2\times$ under adversarial allocation (mudlle and 197.parser, glibc malloc).
Even with modern high-performance allocators, slowdowns up to 27\% are observed.
Clang is the only exception, as not all of its allocations are region-managed,
yielding only partial resistance (3.2--9.4\% depending on the underlying allocator).
Programs with low memory sensitivity, such as 175.vpr and geometry\_nodes,
see little impact from fragmentation regardless of allocator.
\end{tcolorbox}

\section{Threats to Validity}
\label{sec:threats}
This paper's evaluation relies on a specific set of benchmarks that use custom allocators:
we extend a set of five programs taken from Berger et al.'s study~\cite{10.1145/582419.582421}
with three more benchmarks from two modern widely used applications.
These programs are single-threaded; multi-threaded workloads may exhibit
different performance characteristics and memory behavior.
Custom allocators are virtually always used in single-thread or per-thread mode;
therefore comparison in this environment provides a useful baseline.
As discussed in Section~\ref{sec:adv-p1}, these benchmarks also each have a specific allocation size class distribution.
Applications that often use larger allocation sizes such as geometry\_nodes will suffer less
from fragmentation-induced, cache-related performance degradation.

Experiments in this paper are conducted on a server architecture,
detailed in Section~\ref{sec:evaluation}. In particular, our platform has a large
cache size ($450\times$ the last-level cache size of Berger et al.'s processor).
We expect advantages from custom allocators to remain in machines with
equal or smaller cache size, since this reduced capacity will tend to increase the rate of cache misses.
Measurements on a different machine (i7-8559U processor with an 8MB L3 cache) support this claim and
yield results equivalent to those presented in Section~\ref{sec:evaluation}.

Section~\ref{sec:littering} describes adversarial allocation, a stress test for custom and general-purpose
allocation that allows applying increasing degrees of fragmentation. This allows allocator comparison
across a range of behavior, whereas Berger et al.'s previous evaluation inadvertently
focused exclusively on the best-case scenario for the general-purpose allocator. As noted in Section~\ref{sec:precondition},
while this fragmentation may not necessarily occur, at least to that level, during normal execution,
adversarial allocation remains useful to quantify allocator resilience to synthetic fragmentation and simulate tail latency scenarios.

Section~\ref{sec:disable-custom} and Table~\ref{tab:disable-custom} describe the methods used to simulate na\"ive allocation
in each of the benchmarks discussed in this paper.
Switching to na\"ive allocation requires additional object bookkeeping previously handled by the custom allocator;
we follow Berger et al.'s methodology and carefully replace each allocator for maximal performance.
A manual program rewrite moving entirely to explicit memory management may reduce the gap
to region allocation, but will still suffer from diminished contiguity guarantees as discussed in Section~\ref{sec:rq3}.

\section{Related Work}
\label{sec:related}
This paper revisits prior work from Berger et al.~\cite{10.1145/582419.582421}, which classified custom allocator techniques across eight benchmarks.
Section~\ref{sec:evaluation}'s evaluation with modern hardware, allocators, and benchmarks supports the original paper's conclusions.
Namely, only region allocators can provide any significant benefit over na\"ive allocation with a high-performance general-purpose allocator.
We extend our evaluation using adversarial allocation as described in Section~\ref{sec:littering}, surfacing the additional locality guarantees region allocators offer.

Region-based memory management has been studied both as a language construct~\cite{DBLP:journals/iandc/TofteT97} and as an explicit programmer-controlled
interface~\cite{10.1145/277650.277748, 10.1145/378795.378815, 10.1145/512529.512563}, and has more recently been included in the C++ standard
library as of C++17.
The inclusion process produced extensive analysis of region allocator performance and locality benefits~\cite{P0339, P2126, P2127, P2035, N4468, P3002}.
Notably, the N4468 document~\cite{N4468} discusses locality implications of region allocation, using a microbenchmark iterating over an increasingly shuffled vector of subsystems.
We refine this methodology with synthetic fragmentation for any application, allowing evaluation of real programs.

Recent work has sought to bring locality and fragmentation improvements to general-purpose allocators.
In addition to jemalloc~\cite{jemalloc} and mimalloc~\cite{mimalloc} discussed in Section~\ref{sec:evaluation},
Hoard~\cite{hoard} and tcmalloc~\cite{tcmalloc} are other examples of modern high-performance and low-fragmentation general-purpose memory allocators.
\textsc{Mesh}~\cite{10.1145/3314221.3314582} compacts physical memory pages automatically, potentially consolidating objects onto the same cache lines.
However, \textsc{Mesh} requires periodic ``meshing'' and its compaction is probabilistic, hence remaining susceptible to adversarial allocation.
\textsc{Llama}~\cite{10.1145/3373376.3378525} (Learned Lifetime-Aware Memory Allocator) uses machine learning to predict object lifetimes, grouping objects with similar lifetimes together to reduce fragmentation.
While this reduces peak and steady-state memory usage, it adds prediction overhead to each allocation and cannot guarantee spatial contiguity, unlike region allocators which provide it by construction.

\section{Conclusion}
This paper revisits the question of custom memory allocator performance with modern hardware,
general-purpose allocators, and applications.
Berger et al.'s original conclusions hold: per-class allocators provide no substantial benefit
over a state-of-the-art general-purpose allocator, while region allocators remain the only custom
allocation strategy with meaningful performance advantages.
Nonetheless, modern high-performance general-purpose allocators have narrowed the gap to
custom region-based allocators, with region allocator speedups falling from up to 44\% in the original
study to at most 15\% on a clean heap in our evaluation.

We identify a methodological limitation in prior evaluations: starting from a blank heap inadvertently
places general-purpose allocators in ideal conditions, masking a key advantage of region allocators.
Adversarial allocation addresses this limitation by preconditioning the heap with synthetic fragmentation derived
from the program's own allocation behavior.
Under these conditions, the contiguous placement guarantees of region allocators translate directly into
resilience to fragmentation: na\"ive allocation degrades by up to $2\times$, while region-allocated programs are unaffected.
Region allocators are therefore most valuable not just for their raw throughput advantage, but for their
predictability in long-running and tail-latency-sensitive environments.

\begin{acks}
    We thank Joshua Berne, Frank Birbacher, and John Lakos from Bloomberg for many
    productive discussions on this topic and helpful feedback.
    We thank Bloomberg and Meta Platforms for financial support.
    We thank our reviewers and particularly our shepherd, Jeremy Singer, for valuable
    feedback and guidance.
\end{acks}

\bibliographystyle{ACM-Reference-Format}
\bibliography{references}{}
\end{document}